\documentclass{icrc}

\usepackage{times}
\usepackage{graphicx}

\begin{document}

\title{Galactic Magnetic Field Structure and Ultra High Energy Cosmic Ray Propagation}
\author[1,2]{S. O'Neill}
\author[1]{A. Olinto}
\affil[1]{Department of Astronomy and Astrophysics, University of Chicago, Chicago, IL 60637, USA}
\author[3,4]{P. Blasi}
\affil[2]{Department of Astronomy, University of Minnesota, Minneapolis, MN 55455, USA}
\affil[3]{NASA/Fermilab Astrophysics Group, Fermi National Accelerator Laboratory, Batavia, IL 60510-0500, USA}
\affil[4]{Osservatorio Astrofisico di Arcetri, Largo E. Fermi, 5, Firenze, Italy}
\correspondence{smoneil@oddjob.uchicago.edu}

\firstpage{1}
\pubyear{2001}

\maketitle

\begin{abstract}
We consider the effects of the Galactic magnetic field on the propagation of ultra high energy cosmic rays (UHECRs).  By employing two methods of trajectory simulation, we investigate the possibility that UHECRs are produced within the Galaxy with paths strongly influenced by the Galactic magnetic field.  Such trajectories have the potential to reconcile the existing conflict between proposed local sources and isotropic UHECR arrival directions.
\end{abstract}

\section{Introduction}

As is well known, the Greisen-Zatsepin-Kuzmin (GZK) cutoff constrains detected UHECRs to have been produced in or near the Galaxy.  Specifically, detected proton primaries with energies exceeding $5 \times 10^{19}$ eV must have been produced within 50 Mpc of Earth, and nuclei propagation distances are constrained even further \citep{p1976}.  The near-isotropy of detected UHECR arrival directions, however, suggests that Galactic source locations are not easily associated with the observed arrival directions.  We attempt to reconcile these observations by examining the possibility that UHECRs are Galactic in origin, but consist of iron nuclei primaries with trajectories influenced by the Galactic magnetic field.  It has been shown by \citet{b2000} that MHD winds from young neutron stars are capable of Galactic production of iron primaries that fit the observed UHECR energy spectrum.  Such a production mechanism implies that most source locations should be found within the Galactic disk.  With this in mind, we propagate iron nuclei UHECRs through a realistic model of the Galactic magnetic field to investigate the possibility that Galactic sources and isotropic arrival directions are not mutually exclusive phenomena.

\section{Galactic Field Model}

Following \citet{s1997} and \citet{h1999}, we adopt a large-scale regular Galactic magnetic field associated with the spiral arms of the Galaxy.  Specifically, we choose a bisymmetric even-parity field model (BSS-S) in which the field reverses direction between different spiral arms, but is symmetric with respect to the Galactic plane.  The field strength in the plane, directed along the spiral arms, at a point $(\rho,\theta)$ in Galactocentric coordinates is given by
\begin{equation}
B_{sp}=B_{0}(\rho)\cos(\theta-\beta\ln(\rho/\rho_{0}))
\end{equation}
with $\rho_{0}=10.55$ kpc as the Galactocentric distance to maximum field strength at $l=0^{\circ}$ with $\beta=1/\tan{p}$, where the pitch angle is $p=-10^{\circ}$.  The radial dependence of the field strength is given by
\begin{equation}
B_{0}(\rho)=\frac{3r_{0}}{\rho}\tanh^{3}(\frac{\rho}{\rho_{1}})\hspace{2 mm}\mathrm{\mu G}
\end{equation}
where $r_{0}=8.5$ kpc is the Galactocentric distance to the sun and $\rho_{1}=2$ kpc is a smoothing factor that allows for $1/\rho$ behavior beyond 4 kpc from the Galactic center.  The field equation in the Galactic plane is given by
\begin{equation}
\vec{B}(\rho,\theta,z=0)=B_{sp}[\sin p\vec{\hat{\rho}}+\cos p\vec{\hat{\theta}}]
\end{equation}
For this even-parity model, we introduce a $z$ dependence of the following form
\begin{equation}
\vec{B}_{S}(\rho,\theta,z)=\vec{B}(\rho,\theta,z=0)\left(\frac{1}{2\cosh (\frac{z}{z_{1}})}+\frac{1}{2\cosh (\frac{z}{z_{2}})}\right),
\end{equation}
where the values $z_{1}=0.3$ kpc and $z_{2}=4$ kpc reflect the scale heights of the field in the Galactic disk and halo, respectively.  Figure (1) gives a graphical representation of this regular field model, clearly illustrating the field association with the spiral arms.  Note that this model generates field values ranging approximately from $0-6\hspace{1 mm}\mu$G, with a value of $3\hspace{1 mm}\mu $G in the solar neighborhood.    

\begin{figure}[t]
\includegraphics[width=8.3cm]{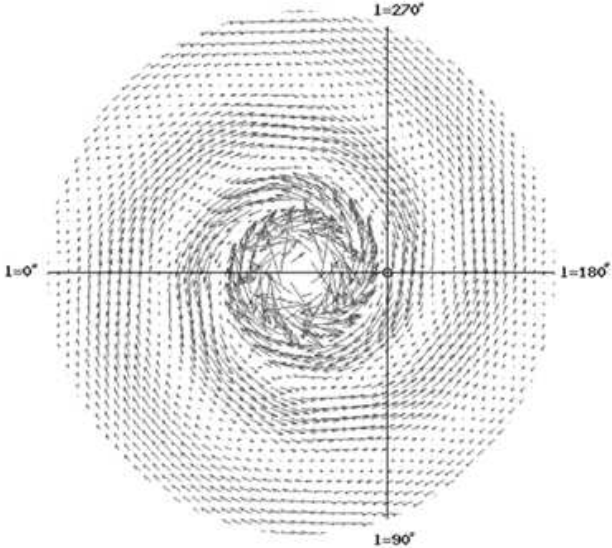}
\caption{Vector plot of the Galactic magnetic field.  The solar position is shown at the intersection of the Galactic coordinate axes.}
\end{figure}

\section{Simulations}

We develop two distinct approaches to the simulation of Galactic UHECR trajectories.  In our first method, we model a distribution of Galactic sources and emit UHECRs according to an assumed $N(E) \propto E^{-1}$ emission energy spectrum in order to study the arrival energy spectrum at a detector.  This allows for comparison between the injected and observed energy spectrum.  Our second approach consists of modeling the antiparticle trajectories as they depart Earth, given the detected UHECR arrival spectrum.  This is equivalent to plotting particle trajectories that are guaranteed to intersect Earth, and this method has the advantage of allowing us to investigate potential source locations corresponding to real observed arrival directions.

\subsection{Particle trajectories}      

Our simulation of UHECR particle trajectories proceeds from a reasonable distribution of neutron stars within the Galactic disk to serve as injection regions for UHECRs.  For simplicity, we assume that all sources are located in the Galactic disk of radius 25 kpc and thickness .65 kpc (taken from the scale height of the thin disk).  We can derive a reasonable number of distinct sources from 
\begin{equation}
N_{ns}=\frac{t_{int}}{r_{ns}}
\end{equation}
where $t_{int}=30$ Myr is the program integration time and $r_{ns}\simeq (1/100\hspace{1 mm}\mathrm{years})$ is the neutron star birth rate.  These values produce $N_{ns}=300,000$ distinct UHECR sources.  Given this set of sources, we assign a random time $t_{emit}<t_{int}$ at which each source emits randomly directed UHECRs according to the $E^{-1}$ energy spectrum, as is expected from neutron star sources \citep{b2000}.  The particles then propagate through the Galaxy with paths influenced by the Lorentz force as they traverse the Galactic magnetic field.

Since an Earth-sized detector is too small to detect a significant number of events with computationally reasonable numbers of injections, we develop a series of larger detectors.  The first is a 2D Galactocentric cylindrical detector of radius $r_{0}=8.5$ kpc, spanning 20-40 pc above the Galactic plane.  The second detector is a 2D Galactocentric ring located 30 pc above the Galactic plane with an inner radius of 8.49 kpc and an outer radius of 8.51 kpc.  These 2D detectors, placed at the solar distance from the Galactic center, possess a number of advantages over local 3D detectors by generating a much higher detection flux while simultaneously reducing the bias for detection of local sources.

With these detectors defined, we sample UHECR energy spectra to compare the detected energy spectrum with the assigned $E^{-1}$ injection energy spectrum.  To accomplish this task, we inject millions of particles from our 300,000 sources and plot the detected energy spectrum.  Figure (2) shows a typical detection energy spectrum from such a simulation.  Since the energy emission spectrum is continuous, we bin the detected data before fitting it to a function of the form $N(E) \propto E^{-\alpha}$.  For the data shown, we calculate $\alpha = 1.00 \pm .09$.  Thus, we find that the emission and arrival spectra are not significantly different for UHECRs.  This is an important result since it is unknown \emph{a priori} if the detection energy spectrum should reflect the emission spectrum when particles have the potential to be trapped in the Galactic field.

\begin{figure}[h]
\vspace*{4.0mm}
\includegraphics[height=8.5cm, width=8.5cm]{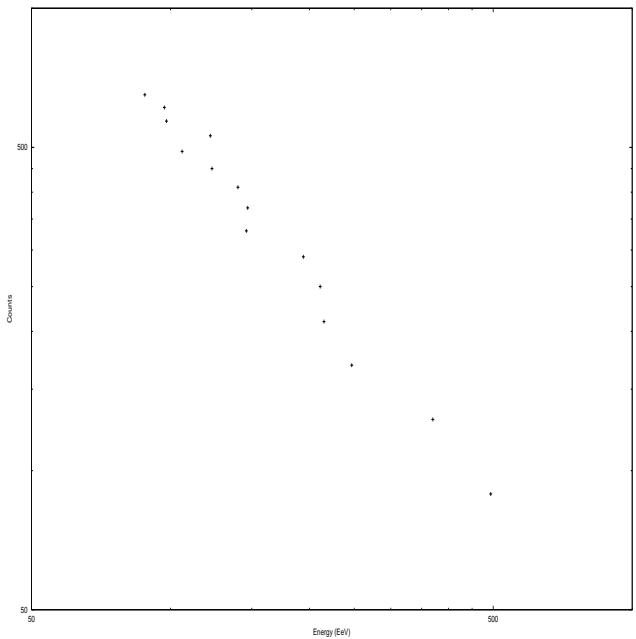}
\caption{Detected energy spectrum from particle simulations.  Fit is of the form $N(E) \propto E^{-1.00 \pm .09}$, matching the emission spectrum.}
\end{figure}

\begin{figure}[h]
\includegraphics[height=8.5cm, width=8.5cm]{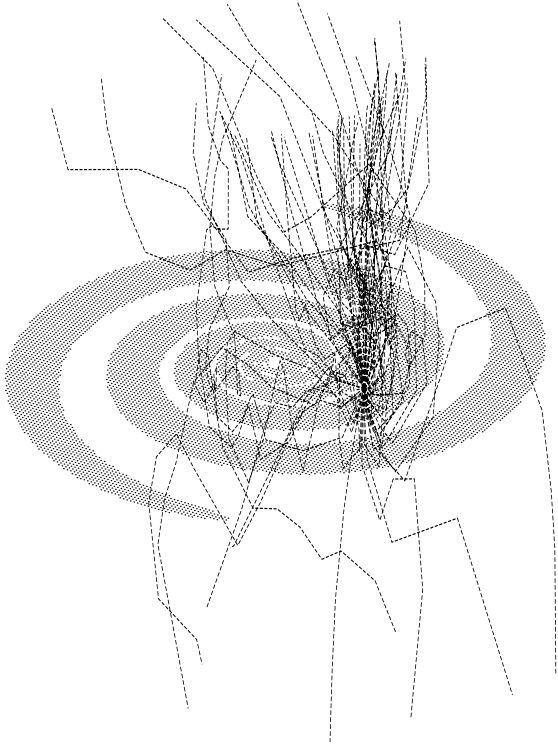}
\caption{50 EeV antiparticle trajectories, emitted from Earth.  The Galactic field is shown in the background for reference.}
\end{figure}   

\subsection{Antiparticle trajectories}

This simulation addresses the other end of the UHECR problem, propagating anti-iron nuclei from a distribution of arrival directions at Earth.  In this approach, we assign energies to the emitted antiparticles and propagate them through the Galaxy to determine possible source locations for the corresponding UHECR particles.  Figure (3) shows one such sample set of trajectories for the case of isotropically distributed 50 EeV iron nuclei.  This method guarantees that our trajectories intersect Earth, but we are now left to evaluate source location without an initial neutron star distribution.  Keeping in mind the need for Galactic neutron stars as sources, we define a potential UHECR source location as any point along an antiparticle trajectory that intersects the Galactic disk.  Furthermore, we closely examine those antiparticle trajectories for which the path first leaves and then reenters the disk.  These particular paths are the least likely to be easily identified with their Galactic sources, and their abundance would lend support to Galactic UHECR origins.  

We proceed by first emitting a known distribution of anti-iron nuclei from Earth.  Specifically, we choose to trace back AGASA events with detected energies that exceed 100 EeV.  Then, using the AGASA energies and arrival directions, we send the particles back through the Galactic field.  Since most of the detected high-energy AGASA events are observed to have arrived away from the plane of the disk, we don't have to worry about the exact difference between arrival direction and source location.  Any path that manages to reintersect the disk some time after emission will have deviated enough to be of interest to us.            

Figure (4) shows the set of trajectories modeled from those AGASA events above 100 EeV.  It is clear that there is little turning back towards the Galactic disk once the particles have exited.  The only path that remains in the Galactic disk for an extended period of time corresponds to the AGASA particle detected nearest the disk.  With our choice of field parameters, the AGASA events do not clearly point back to sources within the Galactic disk.    
\begin{figure}[t]
\includegraphics[height=8.5cm, width=8.5cm]{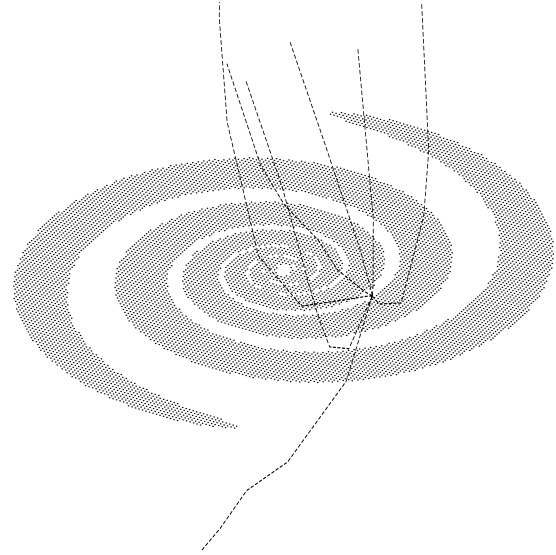}
\caption{Computed trajectories for AGASA events exceeding 100 EeV with a 3 $\mathrm{\mu G}$ Galactic field.}
\end{figure}

\begin{figure*}[t]
\includegraphics[height=8.0cm, width=17.0cm]{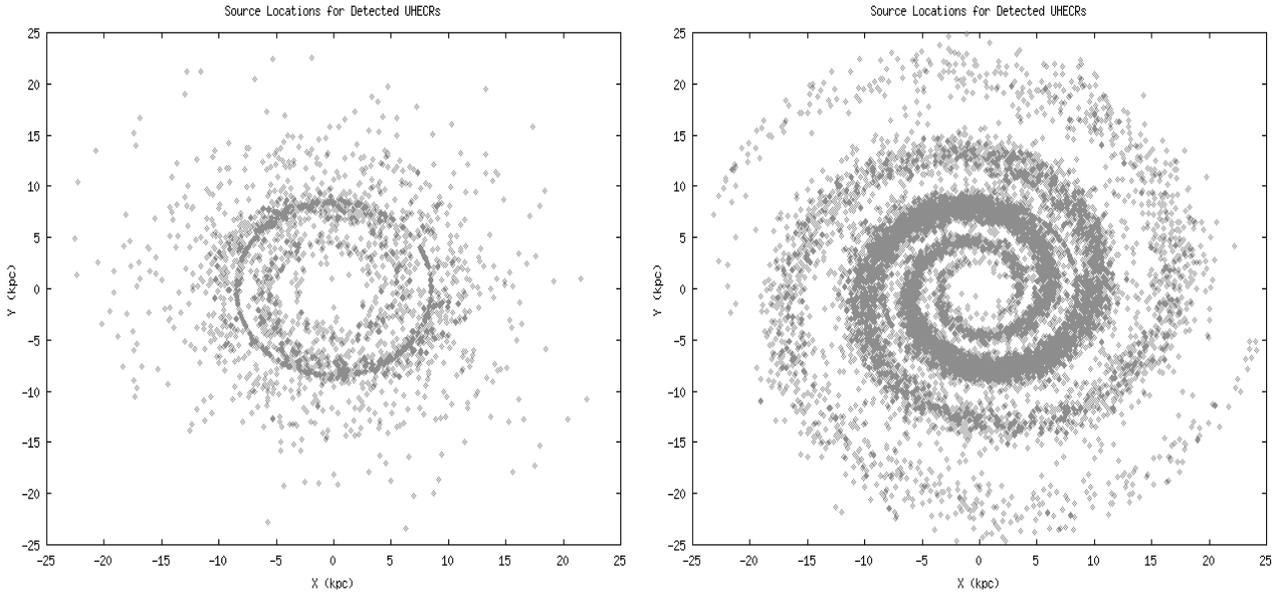}
\caption{Source locations for UHECRs using the Galactic cylindrical detector.  The plot on the left shows locations for a 3 $\mathrm{\mu G}$ field (looking down at the Galactic plane) while the plot on the right is for a 6 $\mathrm{\mu G}$ field.  For reference, the solar system is located at (8.5,0) in each picture.}
\end{figure*}

\section{Discussion and conclusions}

Needless to say, the results of both types of simulation exhibit great sensitivity to certain parameters of the regular magnetic field model.  Specifically, we find that variations in field strength strongly affect possible source locations for detected UHECRs.  Figure (5) shows a contrast between the source location distributions for local field values of 3 and 6 $\mathrm{\mu G}$.  The 3 $\mathrm{\mu G}$ field results in an abundance of detections involving local sources, while the 6 $\mathrm{\mu G}$ case exhibits pronounced non-local Galactic features in the detection spectrum.  Furthermore, differences in the field strength do greatly alter our AGASA trajectories.  Figure (6) shows the trajectories produced from AGASA information with a 6 $\mathrm{\mu G}$ Galactic field.  Clearly, our choice of field parameters is important.

In addition to the regular Galactic field used in our models, there are a number of additional magnetic field components that may be present in the Galaxy.  In particular, a Galactic wind similar in nature to the solar wind can significantly alter UHECR trajectories if the wind field strength is comparable to the regular field strength.  Such a wind model with an azimuthal field strength of 7 $\mu G$ has been proposed to redirect the highest energy particles such that a single source in Virgo can be detected as an isotropic distribution on Earth \citep{a1999, h2000}.  This strong wind effectively funnels isotropic arrival directions to the north Galactic pole for a narrow range of energies.  For iron nuclei, this strong version of the Galactic wind guarantees diffusive behavior up to the highest energies detected thus far.

Local high-field regions within the Galaxy also have the potential to contribute to the diffusion of UHECRs.  Molecular clouds, for instance, are typically associated with mG fields that could serve as scattering regions for cosmic rays.  Our preliminary research has indicated that the known distribution of molecular clouds has a small enough filling factor to cause little change in the paths of most UHECRs, but these and other magnetic field inhomogeneities must be better understood before the picture of UHECR propagation can be made complete.      
\balance
\begin{figure}[h]
\includegraphics[height=8.0cm, width=8.0cm]{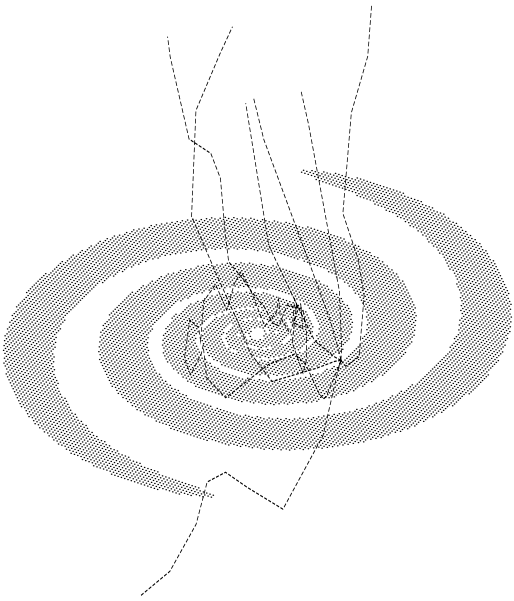}
\caption{Computed trajectories for AGASA events exceeding 100 EeV with a 6 $\mathrm{\mu G}$ Galactic field.}
\end{figure}  

\begin{acknowledgements}
The work of A.V.O and S.O. was supported in part by DOE grant DE-FG0291 ER40606 and NSF grant AST 94-20759.  The work of P.B. was supported by the DOE and NASA grant NAG57092.
\end{acknowledgements}


\begin{thebibliography}{99}

\bibitem[Ahn et al.(1999)]{a1999}
Ahn, E.-J., Medina-Tanco, G., Biermann, P. L., Stanev, T., The origin of the highest energy cosmic rays Do all roads lead back to Virgo?, astro-ph/9911123, 1999.

\bibitem[Blasi et al.(2000)]{b2000}
Blasi, P., Epstein, R. I., and Olinto, A. V., Ultra-high-energy cosmic rays from young neutron star winds, Astrophys. J., 533, L123-L126, 2000.

\bibitem[Harari et al.(1999)]{h1999}
Harari, D., Mollerach, S., and Roulet, E., The toes of the ultra high energy cosmic ray spectrum, astro-ph/9906309, 1999.

\bibitem[Harari et al.(2000)]{h2000}
Harari, D., Mollerach, S., and Roulet, E., Magnetic lensing of extremely high energy cosmic rays in a galactic wind, astro-ph/0005483, 2000.

\bibitem[Olinto(2000)]{o2000}
Olinto, A., Ultra high energy cosmic rays: the theoretical challenge, Phys. Rep., 333-334, 329-348, 2000. 

\bibitem[Puget et al.(1976)]{p1976}
Puget, J. L., Stecker, F. W., and Bredekamp, J. H., Photonuclear interactions of ultrahigh energy cosmic rays and their astrophysical consequences, Astrophys. J., 205, 638-654, 1976.

\bibitem[Stanev(1997)]{s1997}
Stanev, T., Ultra-high-energy cosmic rays and the large-scale structure of the Galactic magnetic field, Astrophys. J., 479, 290-295, 1997.

\end{thebibliography}
\end{document}